\pdfoutput=1
\documentclass[aps,pra,10pt,twocolumn,showpacs,superscriptaddress,floatfix]{revtex4-1}
\usepackage[usenames,dvipsnames,svgnames]{pstricks}

\usepackage{newlfont}
\usepackage{amssymb}
\usepackage{amsfonts}
\usepackage{amsmath}
\usepackage{amsthm}
\usepackage{graphicx}
\usepackage[breaklinks]{hyperref}
\usepackage{color}
\usepackage{bm}
\usepackage{times}

\begin{document}
\title{Quantum Cycle in Relativistic Non-Commutative Space with Generalized Uncertainty Principle correction}

\author{Pritam Chattopadhyay}
\affiliation{Cryptology and Security Research Unit, R.C. Bose Center for Cryptology and Security,\\
Indian Statistical Institute, Kolkata 700108, India}
\email{pritam.cphys@gmail.com}

\author{Tanmoy Pandit}
\affiliation{Hebrew University of Jerusalem, Jerusalem 9190401, Israel}
\email{tanmoypandit163@gmail.com\\ This work was done when the author was visiting Indian Statistical Institute}

\author{Ayan Mitra}
\affiliation{ School of Engineering and Digital Sciences, Nazarbayev University, Nur-Sultan 010000, Kazakhstan}
\email{ayan.mitra@nu.edu.kz}

\author{Goutam Paul}
\email{goutam.paul@isical.ac.in}
\affiliation{Cryptology and Security Research Unit, R.C. Bose Center for Cryptology and Security,\\
Indian Statistical Institute, Kolkata 700108, India}


\begin{abstract}
Quantum heat cycles and quantum refrigerators are analyzed using various quantum systems as their working mediums. For example, to evaluate the efficiency and the work done of the Carnot cycle in the quantum regime, one can consider the harmonic oscillator as it's working medium. For all these well-defined working substances (which are analyzed in commutative space structure), the efficiency of the engine is not up to the mark of the Carnot efficiency. So, one inevitable question arise, can one observe a catalytic effect on the efficiency of the engines and refrigerators when the space structure is changed? In this paper, two different working substance in non-commutative spacetime with relativistic and generalized uncertainty principle corrections has been considered for the analysis of the efficiency of the heat engine cycles. The efficiency of the quantum heat engine gets a boost for higher values of the non-commutative parameter with a harmonic oscillator as the working substance. In the case of the second working medium (one-dimensional infinite potential well), the efficiency shows a constant result in the non-commutative space structure.
\end{abstract}

\maketitle

\section{Introduction} \label{sec1}
Non-commutative (NC) spacetime is an intense area of research in physics. It was the seminal work of Snyder~\cite{hss} which pioneered NC spacetime in Quantum Field Theory, to ameliorate the ultraviolet singularity by incorporating the idea of short-distance cutoff scale via NC spacetime. This work gained its popularity after the work of Seiberg and Witten~\cite{sei}, which specific the existence of low energy limit for open strings with endpoints fixed on D-brans. Even for the analysis and prediction of quantum gravity in low energy phenomena~\cite{das,ali,tka,ghoshs}, new forms of NC spacetime have been developed and explored. Different theories of quantum gravity such as String Theory~\cite{gross,ahar,mague}, Loop Quantum Gravity~\cite{rove,rove1,smolin} and even Doubly-Special Relativity theories~\cite{amelino,mague1} prognosticated the existence of a minimum measurable length (known as Plank length) of spacetime. The existence of the minimum length has a contradiction with the Heisenberg Uncertainty principle~\cite{heisn}, where the latter gives the power to the uncertainty in position to be infinitely small. The description of the existence of the minimum length is given via Generalized Uncertainty Relation (GUP)~\cite{das,kemp,souvik,souvik1,husain,todo}, where the minimum length is considered to be the Planck length, $l_{pl} \, = 10^{-35}m$.

Thermodynamics is a foundation pillar of theoretical physics. It has its impact even in understanding the modern theories like black hole entropy and temperature~\cite{hawk}, gravity~\cite{padma,gur}. It is one of the fundamental theory in the classical regime, but in the quantum regime, the application of thermodynamics need more study. Quantum thermodynamics, in the case of a microscopic system, inspects thermodynamic quantities like temperature, entropy, heat, etc. Even it can analyze the thermodynamic quantities for a single particle model. Exploring heat engines and refrigerators in micro regime~\cite{koslo,skr,quan,jro} is one of the prime focus of the quantum thermodynamics. For the realization of the heat engines in the micro regime, various models are proposed and are even analyzed experimentally~\cite{abah,dech,Zhang}. One can distribute heat engines in two parts, one being discrete in nature, and the other being continuous in its form. In the discrete form, we have two and four-stroke engines, while the turbine is a part of the continuous form. The maximum attainable efficiency is bounded by the Carnot efficiency for the system that is being explored so far. So, the challenge is how we can increase the efficiency of the engines in the micro regimes.  Now, if we change the space structure, how will this affect the analysis of the engines in this space structure.

In this work, we have considered non-commutative spacetime with relativistic and GUP correction for our analysis. Our prime motivation is to inspect the quantum engine cycles for different working mediums in relativistic non-commutative spacetime with GUP correction. Here, the working medium, that we have considered for the analysis is one-dimensional potential well and the harmonic oscillator. We make use of this working medium for the analysis of the Stirling cycle, which depicts the working principle for different quantum heat engines and refrigerators. We evaluate the efficiency of the cycle after the working mediums evolve through every phase of the cycle. The outcomes are surprising in the case of relativistic NC spacetime. We visualize that when the working medium is the one-dimensional potential well, we get a constant efficiency of the engine, whereas, during the case of the harmonic oscillator as the working substance, the efficiency increases rapidly with the variation of the NC parameter of the spacetime. The NC parameter gives a catalytic effect to the efficiency with the harmonic oscillator as the working medium.  This leads us to the possibility of analyzing and exploring quantum information theory from the NC spacetime perspective. One inevitable question that pokes our mind is regarding the physical accessibility of the non-commutative spacetime with the existing quantum technology. In the context of experimental verification and analysis of the signatures of the NC spacetime effect, we encounter recent development in this direction using quantum optics~\cite{sdey} and Opto-mechanical~\cite{mkho} setups.

We have structured the paper as follows. In Section~\ref{sec2}, we analyze the generalized uncertainty relation in the relativistic regime. In Section~\ref{sec3}, we have analyzed the one-dimensional potential problems in the relativistic regime of non-commutative spacetime. Here, we have analyzed two types of potential problems with relativistic and GUP correction, one of which is the harmonic oscillator, and the other is the one-dimensional infinite potential well. We have devoted Section~\ref{sec4} to develop the Stirling cycle with the harmonic oscillator and one-dimensional potential well with relativistic and GUP correction as the working medium. In this section, we have demonstrated the work done and the efficiency of the quantum engine. We have concluded the paper in Section~\ref{sec5} with some discussions regarding the analysis done in this paper and its future prospects in the field of quantum thermodynamics.


\section{Generalized Uncertainty Principle in relativistic regime}\label{sec2}
Here, we revisit the analysis of generalized uncertainty principle~\cite{todo} in the relativistic regime.
The well known quadratic GUP was first proposed in the work~\cite{kemp}, which takes the form as
\begin{equation}\label{a1}
    [x^i, p^j] = i \hbar \delta^{ij} (1+g(p^2)),
\end{equation}
where $i,j \,\, \epsilon \,\, \{1,2,3\}$, with $\delta^{ij}$ being the Kronecker delta function which results to $1$ when $i=j$ and zero otherwise. Here, $x$ and $p$ represent the position and momentum of a particle respectively. This form of the space structure was proposed for the non-relativistic regime. The position operators for this model obeys the commutation relation 

\begin{equation}\label{a2}
    [x^i,x^j] = -i \hbar g(p^2) (x^i p^j - x^j p^i).
\end{equation}

For our analysis we consider a non-commutative space~\cite{todo} which obeys the following commutation relation 

\begin{equation}\label{a3}
    [x^i, p^j] = i \hbar \Big(\Big[1+(\epsilon - \alpha) \, \zeta^2  p^\delta p_\delta \Big] \eta^{ij} \, + (\beta + 2\gamma)\, \zeta^2 p^i p^j \Big),
\end{equation}
where  $i,j \,\, \epsilon \,\, \{0,1,2,3\}$ and $\alpha$, $\epsilon$, $\beta$, $\gamma$ are dimensionless parameters. The parameter $\zeta$ takes the form $\zeta = \frac{1}{c\, M_{pl}}$ and has the dimension of inverse momentum and $\eta^{ij}$ takes the signatures $\{-, + , +, + \}$ of the Minkowski spacetime. Here, $M_{pl}$ is the Planck mass. Eq.~\eqref{a3} reduces to non-relativistic limit (Eq.~\eqref{a1}) when $c\rightarrow \infty$, and when $\zeta \rightarrow 0$ the system boils down to the non-GUP limit where the standard Heisenberg algebra works.  

If we take a clear note, we can visualize that the physical observables (the position and the momentum) of the system are not canonically conjugate. By introducing the variables $x_0^i$ and $p_0^i$ (where $p_0^i = - i \hbar \frac{\partial}{\partial x_{0i}}$) which are canonically conjugate in nature the position and the momentum can be expressed up to the second-order of $\zeta$ as

\begin{eqnarray}\label{a4}\nonumber
    x^i & = & x_0^i - \alpha \zeta^2 p_0^\delta p_{0 \delta} x_0^i + \beta \zeta^2 p_0^i p_0^\delta x_{0\delta} + \gamma \zeta^2 p_0^i,\\
    p^i & = & p_0^i \,(1+\epsilon \zeta^2 p_0^\delta p_{0 \delta}).
\end{eqnarray}
Using Eq.~\eqref{a4} the commutation relation for the position operators becomes 
\begin{equation}\label{a5} 
    [x^i, x^j] = i \hbar \zeta^2 \frac{2\alpha + \beta}{1+(\epsilon - \alpha) \, \zeta^2  p^\delta p_\delta} (x^i p^j- x^j p^i).
\end{equation}

The last two terms in the expression of $x^i$ of Eq.~\eqref{a4} break the isotropy of  the spacetime and violates the relativity principles while introducing the preferred direction of $p_0^i$. So, from now onward we will consider $\beta = \gamma = 0$ for further analysis. 


\section{One-dimensional potential problems in NC space with relativistic and GUP correction}\label{sec3}

In this section, we are going to revisit the analysis of some of the one-dimensional (1-D) potential systems for this non-commutative space structure. For the analysis, we substitute $\epsilon = \alpha$ in Eq.~\eqref{a4} (we use this condition to keep the Poincare algebra undeformed) and neglecting the last two terms of $x^i$ we have
\begin{eqnarray}\label{b1}
    x^i & = & x_0^i - \alpha \zeta^2 p_0^\delta p_{0 \delta} x_0^i,\\ \nonumber
    p^i & = & p_0^i \,(1+\alpha \zeta^2 p_0^\delta p_{0 \delta}).
\end{eqnarray}

The Klein-Gordon (KG) equation in terms of the variables $p_0^i$ is

\begin{equation}\label{b2}
    p_0^\delta p_{0 \delta} (1 + 2 \alpha \zeta^2 p_0^\rho p_{0 \rho}) = - m^2 c^2,
\end{equation}
where $m$ is the mass of the relativistic particle. Solving Eq.~\eqref{b2} in terms of $p_0^\delta p_{0 \delta}$ we get
\begin{eqnarray}\label{b3} \nonumber
   p_0^\delta p_{0 \delta} & = & - \frac{1}{4\alpha\zeta^2} + \Big(\frac{1}{(4\alpha\zeta^2)^2} -\frac{m^2 c^2}{2 \alpha \zeta^2} \Big)^{\frac{1}{2}},\\
   & \simeq & - m^2 c^2 - 2 \alpha \zeta^2 m^4 c^4 - \mathcal{O} (\zeta^4).
\end{eqnarray} 
The higher-order terms are discarded, and along with that, the other solution of the KG equation is not taken into account as it does not reduce to $m^2 c^2$ when $\zeta \rightarrow 0$.

One can rewrite Eq.~\eqref{b3} as 
\begin{equation} \label{b4}
    -E^2 + c^2 p_0^2 +m^2 c^4 + 2 \alpha \zeta^2 m^4 c^6 = 0.
\end{equation}
Now solving the Eq.~\eqref{b4}, we can evaluate the expression of the energy for the system. It takes the form
\begin{eqnarray} \label{b5} \nonumber
    E& = & mc^2 (1+\alpha \zeta^2 m^2 c^2) + \frac{p_0^2}{2m} \Big(1-\frac{1}{2} \alpha \zeta^2 m^2 c^2\Big)\\
    & - & \frac{p_0^4}{8m^3c^2} \Big(1-3\alpha \zeta^2 m^2 c^2 \Big).
\end{eqnarray}
The energy expression consists of the rest mass term, the non-relativistic kinetic energy term, along with that it possesses relativistic and GUP corrections. Now, the  Schr\"odinger equation with relativistic and the GUP corrections, is defined as 

\begin{eqnarray}\label{b6} \nonumber
i\hbar \frac{\partial}{\partial t_0} \psi(t_0,x_0) & = & \Big[mc^2(1 + \alpha \zeta^2 m^2 c^2) \\ \nonumber
& + & \frac{\hbar^2}{2m} (1-\frac{1}{2} \alpha \zeta^2 m^2 c^2) \nabla_0^2 \\ \nonumber
& + & \frac{\hbar^4}{8m^3c^2} (1-3\alpha \zeta^2 m^2 c^2) \nabla_0^4 \\
& + & V(x) \Big] \psi(t_0,x_0).
\end{eqnarray}
    
One can solve Eq.~\eqref{b6} for a different potential problem with relativistic and GUP correction, to develop the wavefunction and the physical energy of the considered problem. We will consider two such potential problems for our analysis. One of which is the one-dimensional infinite potential well and the other is the harmonic oscillator.

\subsection{One Dimensional Potential Well}
The 1-D potential well for 1+1 dimensional case is defined as 
\begin{eqnarray}\label{c1}
V(x)=\left\{
\begin{array}{lc}
  V_0, &\ for \ \ 0 < x < L ,\\
   \infty, & \ for\ \  x \leq {0} \ \cup \ x \geq {L}.
\end{array}\right.
\end{eqnarray}
The physical dimensions of the 1-D box for the system under consideration can be evaluated from Eq.~\eqref{b1}. It s expressed as 
\begin{equation} \label{c2}
    L= L_0 [1 + \alpha \zeta^2 m^2c^2 + \mathcal{O}(\zeta^4)].
\end{equation}
Solving the Schr\"odinger equation (Eq.~\eqref{b6}) for this potential problem, the energy of the system results to 
\begin{equation}\label{c3}
    E_n = -\frac{n^2\hbar^2\pi^2}{2mL^2} \Big[1 + \frac{3}{2} \alpha \zeta^2 m^2 c^2 \Big] - \frac{\hbar^4}{8m^2 c^2} \Big[\frac{n\pi}{L}\Big]^4.
\end{equation}
Here, the first term in the Eq.~\eqref{c3} corresponds to the non-relativistic energy of the system with GUP-corrections, whereas, the last term of the expression depicts the relativistic corrections.

\subsection{Harmonic Oscillator}
Harmonic oscillator (HO) is a well defined potential problem. Here we are going to analyze the harmonic oscillator for the system under consideration in a 1+1 dimensional case. The harmonic oscillator potential is 
\begin{eqnarray}\label{d1}\nonumber
V(x)= \frac{1}{2} m \omega^2 x^2.
\end{eqnarray}
Using this potential in the Schr\"odinger equation (Eq.~\eqref{b6}) and solving it, we can evaluate the expression of the energy for this non-commutative spacetime model. The expression of the energy is
\begin{eqnarray}\label{d2} \nonumber
    E_n & = & \hbar \omega \Big(n + \frac{1}{2} \Big) \Big[1- \frac{1}{2} \alpha \zeta^2 m^2 c^2 \Big] \\
        & - & \frac{\hbar^2 \omega^2}{32 m c^2} \Big[1 - 4 \alpha \zeta^2 m^2 c^2\Big] \Big(5n(n+1) + 3\Big). 
\end{eqnarray}
If one calculates the landau levels similar to the previously defined work~\cite{das}, one will come up with a bound on $\alpha$. The bound is
\begin{equation*}
    \alpha \leq 10^{41}.
\end{equation*}
    
Hereafter, for our analysis, we will consider $\alpha \thicksim 10^{41}$.


\section{Quantum heat cycles in NC with GUP corrections} \label{sec4}
Our new results are presented in this section. Quantum heat engines and refrigerators are analyzed for different systems as the working substance. We will consider the 1-D potential well with relativistic and GUP correction as the working substance for the engine model. We will also take into account the harmonic oscillator (HO) with relativistic and GUP correction as the working substance for the engine model. Our purpose is to analyze the better working medium for the heat engine cycles in non-commutative space in the relativistic realm. For our analysis, we will study the Stirling engine cycle for both the cases. In the non-commutative phase space, the working medium that we will consider for the analysis of the Stirling cycle will evolve to a Gibbs state which we will consider for our analysis following the same analogy as shown in previous works~\cite{c1234}.

\begin{figure}[ht]
  \centering
  \includegraphics[width=1.0\columnwidth]{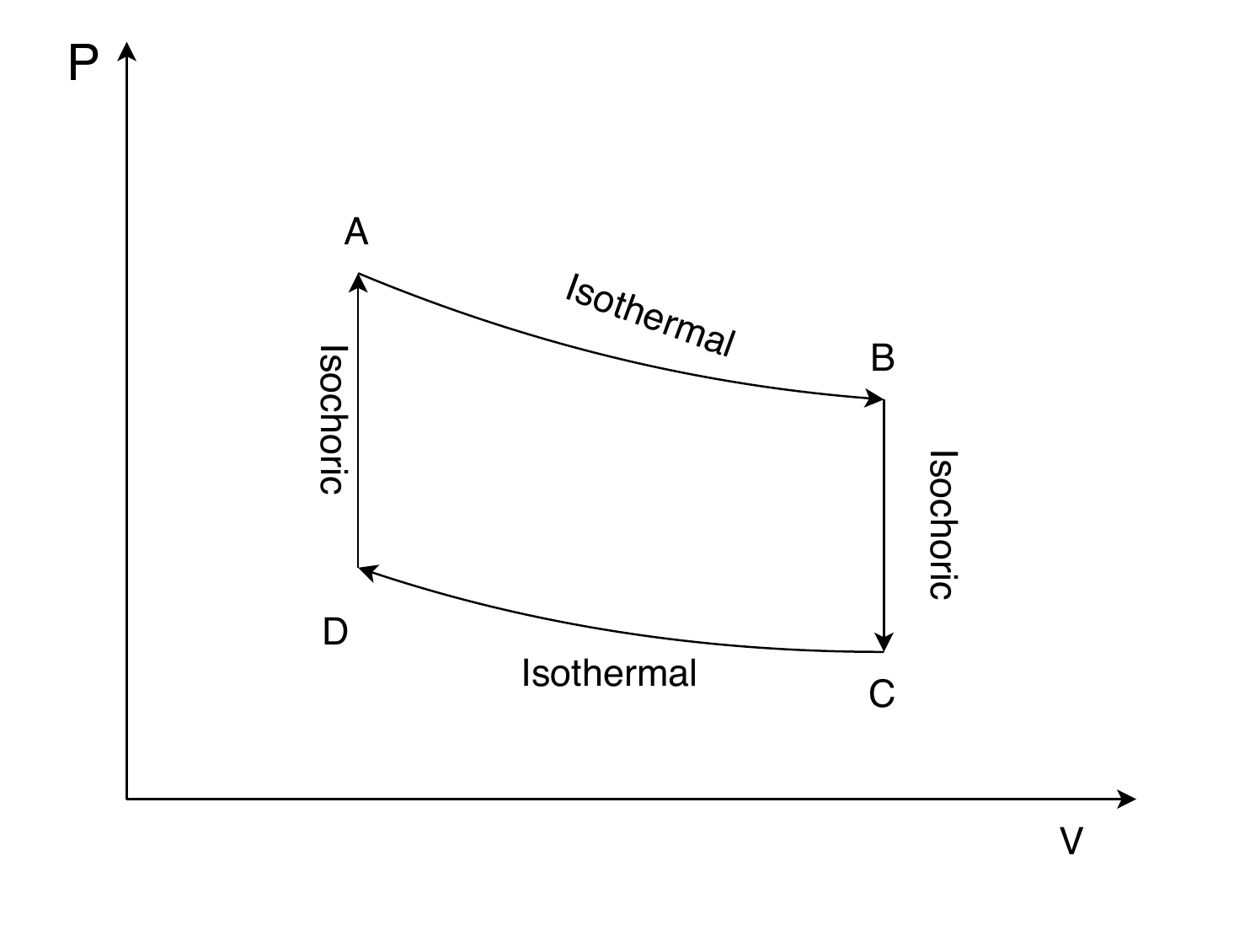}
\caption{Schematic representation for the $4$-stages of the Stirling cycle.}
\label{Fig1}
\end{figure}

Stirling cycle, a reversible thermodynamic cycle, is a four-stroke engine that comprises of two isothermal processes and two isochoric processes. The pressure-volume (P-V) diagram of the Stirling cycle in the classical regime is depicted in Fig.~(\ref{Fig1}).

In Fig.~\ref{Fig1}, $AB$ and $CD$ represents the isothermal processes. $BC$ and $DA$ describes the isochoric process in the cycle. The system is coupled with the hot bath during the $DA$ and $AB$ phase of the cycle. The system gets coupled with the cold bath during $BC$ and $CD$ phase of the cycle. Work done occurs only during the isothermal processes.

\subsection{Stirling cycle with 1-D well as working substance}
Here, we analyze the Stirling cycle with 1-D potential well as the working substance. A Stirling cycle~\cite{say,agar,hua,blick} consists of four stages, two of which are isothermal processes, and the remaining two processes are isochoric in nature.

(i) First stage of the Stirling cycle is the \textit{isothermal (A$\rightarrow$B)} process. In this phase, we insert a barrier isothermally in the 1-D well, which divides the well into two equal halves. The positioning of the partition (i.e. barrier) in the middle of the infinite 1-D well converts it to an infinite 1-D double potential well. In our analysis, we have taken into account that the delta potential grows in strength from zero to that height, which would be enough to prevent tunneling process through the barrier. This provides us the assurance that the probability tends to zero for the tunneling process considering that the time required for the execution of the tunneling process is more than the time required for any thermodynamic processes to complete. During this phase of the cycle, the working medium is attached to a hot bath with a temperature $T_1$. Throughout this process, the system remains at its equilibrium condition when the quasi-static insertion of the barrier is being done.

(ii) The second phase: \textit{isochoric (B$\rightarrow$C)} process. In this stage of the cycle the working medium experiences an isochoric heat extraction. The system is connected to the cold bath at temperature $T_2$ where $T_1 > T_2$.

(iii) The third phase: \textit{isothermal (C$\rightarrow$D)} process. In this stage of the cycle, we remove the barrier from the 1-D well isothermally. The execution of this stage is carried out keeping in mind that the tunneling probability tends to zero, and it remains in equilibrium at temperature $T_2$.

\begin{figure}
  \centering
  \includegraphics[width=1.0\columnwidth]{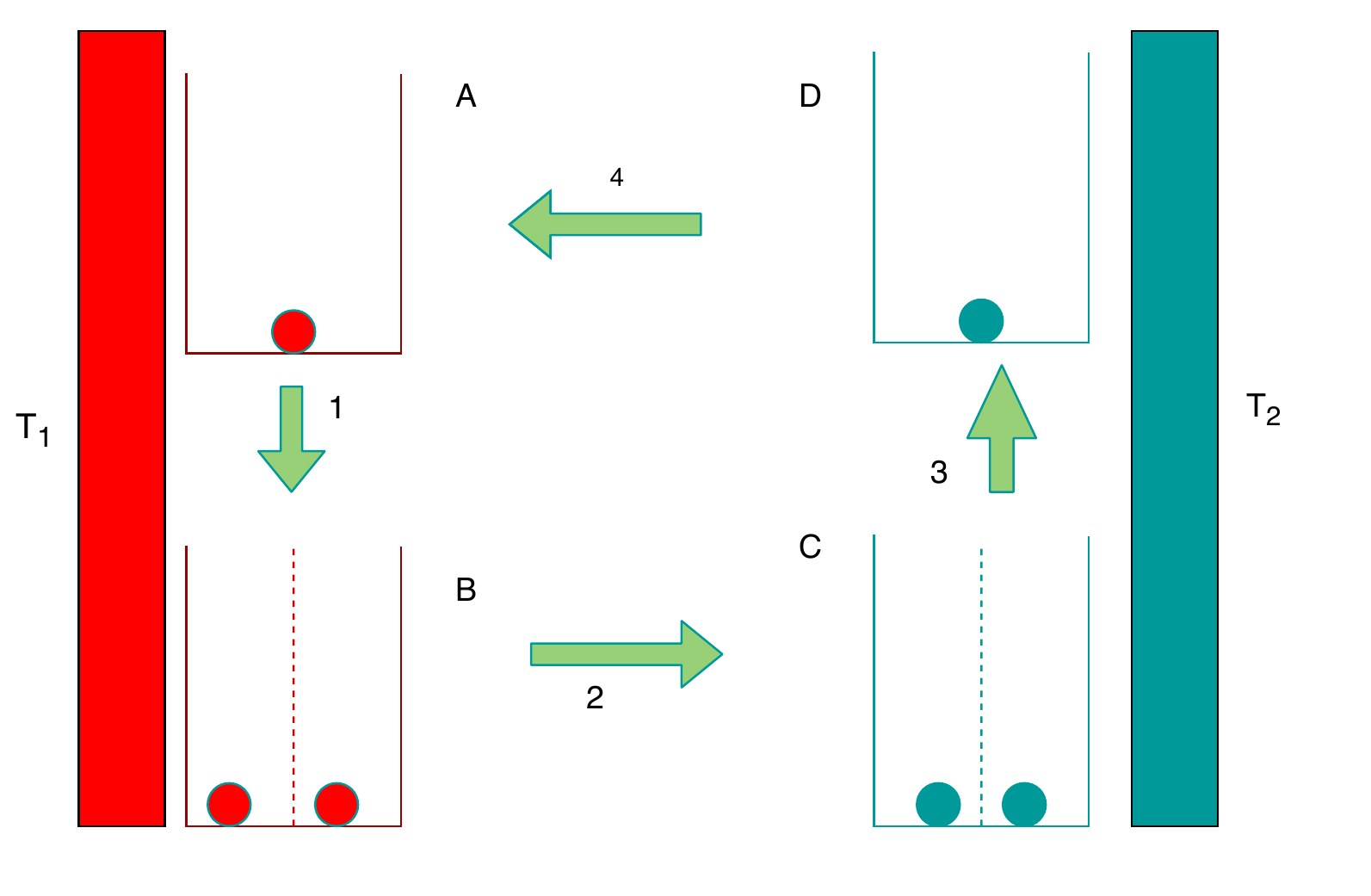}
\caption{Schematic representation for the $4$-stages of the Stirling cycle with 1-D potential well as working substance}
\label{Fig2}
\end{figure}

(iv) The fourth stage:  \textit{isochoric (D$\rightarrow$A)} process. During this last phase of the cycle, isochoric heat absorption is observed when the working medium is connected back to the bath at a temperature $T_1$. The schematic representation of the cycle is shown in Fig.~(\ref{Fig2}).

In previous works~\cite{thomas,chatt,chatt1,ssingh}, they have studied the work done and the efficiency for the heat engine in the non-relativistic and relativistic regime in commutative space. In this work, we first develop the heat engine in the NC spacetime with relativistic and GUP correction with one-dimensional potential well as the working medium. Following the same analogy in the previous works, we analyze the work done and the efficiency of the heat engine for NC spacetime.

\textcolor{black}{Some previous works~\cite{b12,b13} have explored the non-equilibrium thermodynamics in non-commutative space structure. They have shown that the master equation for the non-commutative space narrow down to the usual master equation when one equates the non-commutative parameter to zero. The presence of the non-commutative parameter enhances the results that are generally encountered in the ordinary space without violating the usual master equation. We have considered that our system follows the same usual master equation in the non-commutative space as suggested in the works~\cite{b12,b13}. In our analysis we have considered that, only the system will encounters the effect of the non-commutative space.}

Now, we will take a 1-D potential well having a length $2L$  in NC spacetime with a particle inside the system having mass $m$ at temperature $T_1$ as the working medium for our analysis. The physical energy for this system is identical to Eq.~(\ref{c3}). 

The partition function~\cite{reif} of our system is 
\begin{eqnarray}\label{e1} \nonumber
   Z_A  & = & \sum_n e^{-\beta_1 E_{n}}\\
  & = & \frac{1}{\sqrt{\pi} \sqrt{\frac{\beta_1 \hbar^2(2 + 3 \alpha \zeta^2 m^2 c^2)}{L^2 m}}}, 
\end{eqnarray}
where $\beta_1 = \frac{1}{k_B T_1}$ and $k_B$ is the Boltzmann constant.

Now, in the first stage of the cycle, we divide the 1-D well into a double potential well by inserting a partition isothermally. During this process, even energy levels remain therein, but we observe a shift in the odd levels. The odd ones merge with the nearest neighbor, i.e., the even energy level. So, the energy after the insertion takes the form 
 
\begin{equation}\label{e2}
E_{2n} = -\frac{{(2n)}^2\hbar^2\pi^2}{8mL^2} \Big[1 + \frac{3}{2} \alpha \zeta^2 m^2 c^2 \Big] - \frac{\hbar^4}{8m^2 c^2} \Big[\frac{(2n)\pi}{2L}\Big]^4,
\end{equation}
which evolves by substituting $n$ by $2n$ in Eq.~\eqref{c3}. The partition function after the insertion of the well is defined  equivalent to Eq.~\eqref{e1} as
\begin{eqnarray}\label{e3} \nonumber
Z_B = \sum_n 2e^{-\beta_1 E_{2n}}. 
\end{eqnarray}

 In the isothermal process, the heat exchange takes the form  
\begin{equation}\label{e4}
Q_{AB} = U_B - U_A + k_B T_1 ln Z_B - k_B T_1 ln Z_A.
\end{equation}

 The internal energy $U_A$ and $U_B$ is described as $U_i = - \partial ln Z_i \big/ \partial \beta_1$, where $i=A,B$.
 
During the second stage of the cycle, the system is connected to the cold bath at a temperature $T_2$. The partition function takes the form 
\begin{equation}\label{e5} \nonumber
Z_C = \sum_n 2e^{-\beta_2 E_{2n}}.
\end{equation}

The heat exchanged during this stage of this cycle is interpreted by the difference of the average energies for the initial and final states. It takes the form  
\begin{equation}\label{e6} 
Q_{CB} = U_C - U_B,
\end{equation}
where $U_C = - \partial ln Z_C\big/ \partial \beta_2$ and $\beta_2= \frac{1}{k_B T_2}$. In the third phase of the cycle, we remove the wall isothermally. The energy during this stage of the cycle reverts back and is the same as given in Eq.~\eqref{c3}. The  partition function  becomes 
\begin{equation}\label{e7} \nonumber
Z_D = \sum_n e^{-\beta_2 E_n},
\end{equation}
where  $U_D$ can be evaluated similarly as $U_C$. The heat exchanged  (similar to Eq.~\eqref{e4}) becomes
\begin{equation}\label{e8}
Q_{CD} = U_D - U_C + k_B T_2 ln Z_D - k_B T_2 ln Z_C.
\end{equation}

In the fourth stage of the cycle, the system falls back into the first phase of the cycle, i.e., the system is bridged back to the heat reservoir at  temperature $T_1$. The  heat exchange for the system is expressed as 
\begin{equation}\label{e9}
Q_{DA} = U_A- U_D.
\end{equation}

The total work done for this cycle is evaluated using Eq.~\eqref{e4}, ~\eqref{e6}, ~\eqref{e8} and ~\eqref{e9} as
\begin{eqnarray}\label{e10} \nonumber
W & = & Q_{AB} + Q_{BC} + Q_{CD} + Q_{DA}.
\end{eqnarray}

The efficiency of this engine  is defined using Eq.~\eqref{e4}, ~\eqref{e6}, ~\eqref{e8} and ~\eqref{e9} as

\begin{eqnarray} \label{e11}\nonumber
\eta_{Stir} & = & 1 + \frac{Q_{BC} + Q_{CD}}{Q_{DA} + Q_{AB}}.
\end{eqnarray}

\begin{figure}[h]
  \centering
  \includegraphics[width=1.0\columnwidth]{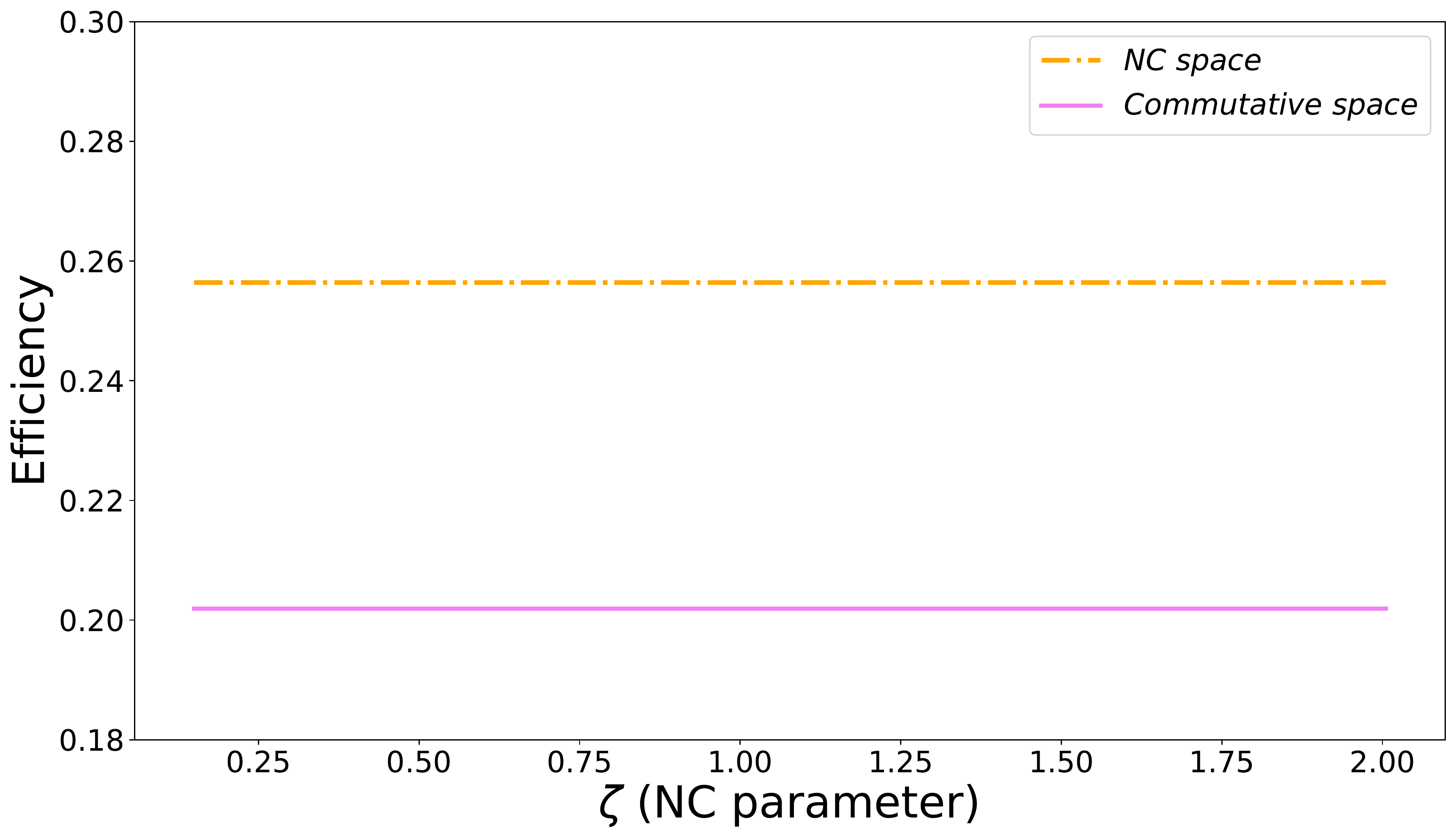}
\caption{(Color online) Variation of the efficiency of Stirling cycle with the NC parameter of the system with 1-D potential as the working substance. The violet solid line depicts the efficiency of the cycle in commutative phase space and the orange dashed-dot curve represents the efficiency in non-commutative phase space.}
\label{Fig3}
\end{figure}

We have considered $T_1 = 2K$ and $T_2 = 1K$ for the evaluation the efficiency of the Stirling engine (shown in Fig.~(\ref{Fig3})) with the variation of the NC parameter. The efficiency of the engine has been numerically analyzed with the results of the partition function and the internal energy of the system. We have considered the length of the potential well as $5 \, nm$. The efficiency of the engine is constant with the variation of the parameter. So, for this working medium, we do not encounter any boost in the efficiency of the cycle. If we analyze the engine cycle with this working medium in commutative space, the efficiency results to a constant value which is near about $(0.2)$ for a fixed length of the potential well. Along with that, we encounter that the efficiency of the engine is not up to the mark with the relativistic and GUP correction working medium. We can infer from the plot shown in Fig.~(\ref{Fig3}) that the non-commutative parameter has lost its impact when we have considered this model as the working medium. The gain in the efficiency of he engine is provided by the relativistic correction. This is quite surprising as we have encountered gain due to the non-commutative parameters as shown in previous works~\cite{a12,a13,a14,a15}. This feature of suppressing the non-commutative affect with relativistic correction needs further investigation.

\subsection{Stirling cycle with harmonic oscillator as working substance}
\begin{figure}
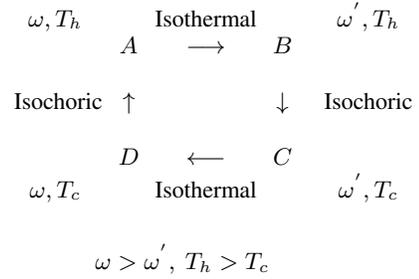

\begin{align}\nonumber
\label{1a}
&\begin{array}{cccccc}
\omega,T_h~&  &\text{Isothermal}& &&\omega^{'},T_h\\
& A &\longrightarrow& B&&\\
&  &&\\
 \text{Isochoric}&\uparrow&&\downarrow&&\text{Isochoric}\\
&  &&\\
& D&\longleftarrow&C &&\\
\omega,T_c~& &\text{Isothermal}& &&\omega^{'},T_c\\
&&&&&\\ \end{array} \\ \nonumber
&~~~~~~~~~~~~~~~\omega>\omega^{'},~T_h>T_c~~~~~~~~~~\\ \nonumber
\nonumber
\end{align}
\caption{Stirling cycle with harmonic oscillator as working substance}
\label{1aa}
\end{figure}

Now, we analyze the Stirling cycle having a harmonic oscillator with relativistic and GUP correction as the working substance. A Stirling cycle~\cite{say,agar,hua,blick}, consists of four stages, two isothermal processes, and the remaining two processes are isochoric. The schematic representation of the Stirling engine cycle with HO as a working substance is shown in Fig.~\ref{1aa}.

The physical energy of the system is equivalent to Eq.~\eqref{d2}. Now, the partition function~\cite{reif} for the system is defined as 
\begin{eqnarray} \label{f1} \nonumber
Z & = & \sum_n e^{-\beta E_{n}} \\ 
& =& \frac{2 (\frac{2\pi}{5})^\frac{1}{2}  e^\kappa \chi}{\Theta},
\end{eqnarray}
where $\kappa = \frac{\beta \Xi }{640 c^2 m  \Big(-1 + 4 c^2 m^2 \alpha \zeta^2 \Big)}$, $\Xi = \Big(-1024c^6 m^4 \alpha \zeta^2 + 256 m^6 \alpha^2 \zeta^4 - 35 \hbar^2\omega^2 +280 c^2\hbar^2 m^2 \alpha \zeta^2 \omega^2 - 16c^4 m^2 \Big(-64 + 35 \hbar^2 m^2\alpha^2\zeta^4 \omega^2  \Big)\Big)$, $\Theta = \sqrt{\beta \omega^2 (4\alpha \zeta^2-1)}$ and $\chi = 1+ erf \Big[\frac{\hbar \beta \omega \Big(16c^4m^3\alpha\zeta^2 + 5 \hbar \omega - 4 c^2m\Big(8 + 5 \hbar m \alpha \zeta^2 \omega \Big)\Big)}{\Theta} \Big]$. Here $erf$ is the error function~\cite{abcd} which is defined as $erf(x) = \pi^{-\frac{1}{2}} \Gamma(\frac{1}{2}, x^2)$, where $\Gamma (\alpha, x)$ is known as incomplete gamma function~\cite{amore}.

(i) The first stage: \textit{isothermal (A$\rightarrow$B)} process. During this phase, the working medium is attached with a heat reservoir of temperature $T_h$. Throughout this process, the system will be in thermal equilibrium with the heat bath. The changes in the energy spectrum and the internal energy of the system take place as a result of quasistatic changes that occurs to the working medium. These changes are due to the changes that occur in the Hamiltonian of the system when the system evolves during this process. The heat exchange that occurs during the first stage of the Stirling cycle is
\begin{eqnarray} \label{f2}
Q_{AB} = U_B - U_A + k_B T_h ln Z_B - k_B T_h ln Z_A,
\end{eqnarray}
where $k_B$ is the Boltzmann constant. The partition function $Z_A$, $Z_B$ for the phases can be calculated using Eq.~\eqref{f1}.  The internal energy $U_A$ and $U_B$ is described as $U_i \equiv - \partial ln Z_i \big/ \partial \beta_h$, where $i=A,B$.

(ii) The second phase: \textit{isochoric (B$\rightarrow$C)} process. During this phase, the system evolves under an isochoric heat exchange. The system is coupled with a bath at a temperature $T_c$, so heat will be released in this process. So, the heat exchange is expressed as 

\begin{eqnarray} \label{f3}
Q_{BC}= U_C -U_B.
\end{eqnarray}

(iii) Third phase: \textit{isothermal (C$\rightarrow$D)} process. In this process, the working substance remains coupled with the bath at temperature $T_c$. Similar to the first isothermal process, the system during this process is at thermal equilibrium with the bath. Heat gets released in this phase. The heat exchange takes the form  

\begin{eqnarray} \label{f4}
Q_{CD} = U_D - U_C + k_B T_c ln Z_D - k_B T_c ln Z_C.
\end{eqnarray}
where $U_C= - \partial ln Z_C\big/ \partial \beta_c$ and $\beta_c= \frac{1}{k_BT_c}$. 

(iv) The fourth phase: \textit{isochoric (D$\rightarrow$A)} process. The system falls back into the bath with  temperature $T_h$. The heat exchange for this stage is expressed as
\begin{eqnarray} \label{f5}
Q_{DA} = U_A-U_D.
\end{eqnarray}

The total work done for this process is $W_{tot} = Q_{AB} + Q_{BC} + Q_{CD} + Q_{DA}$. The efficiency of the Stirling heat cycle from Eq.~\eqref{f2}, ~\eqref{f3}, ~\eqref{f4} and ~\eqref{f5} is expressed as 

\begin{eqnarray} \label{f6} \nonumber
\eta_{Stir}= 1 + \frac{Q_{BC} + Q_{CD}}{Q_{DA} + Q_{AB}}.
\end{eqnarray}

\begin{figure}
  \centering
  \includegraphics[width=1.0\columnwidth]{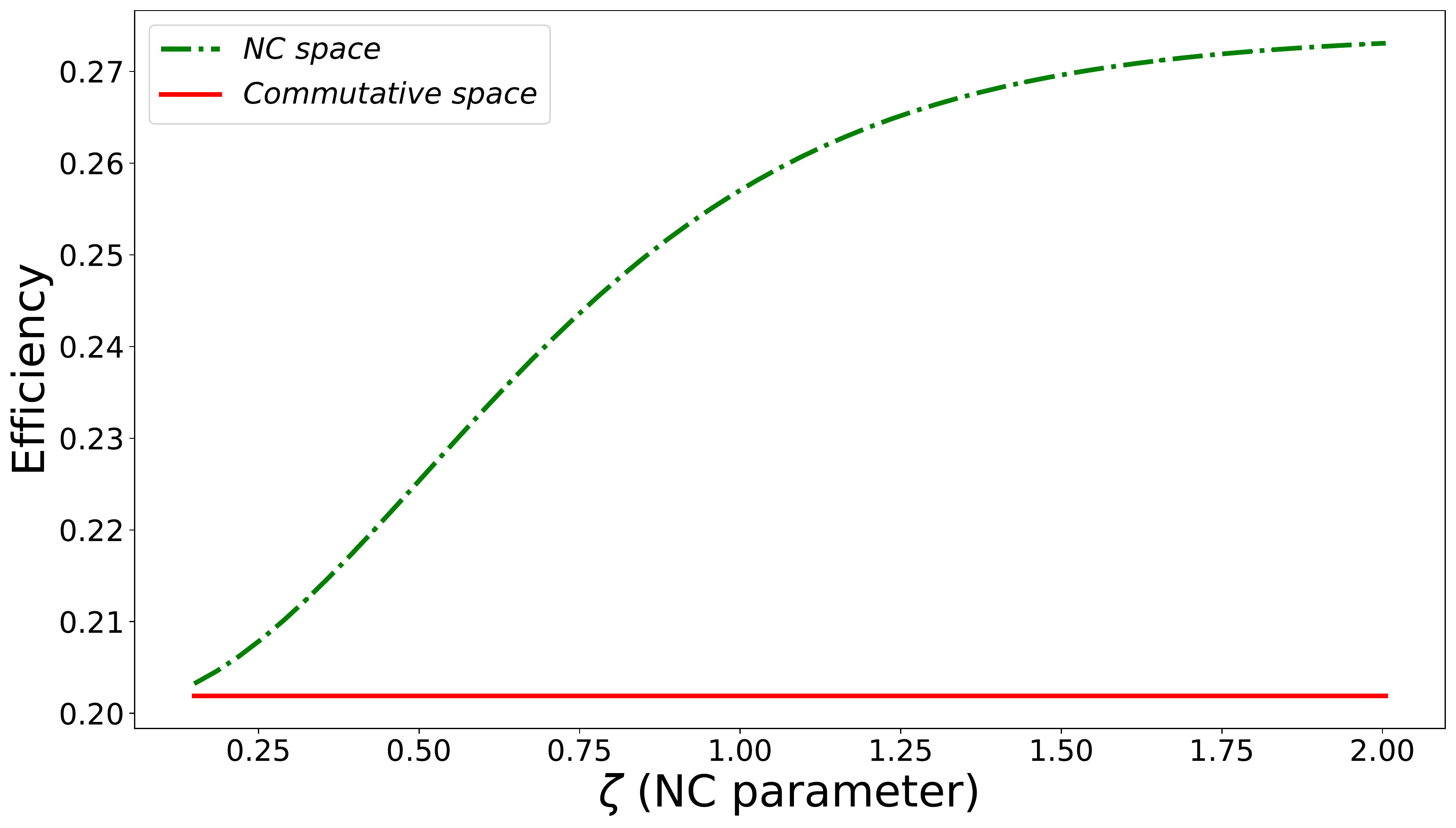}
\caption{(Color online) The green dash-dot curve shows the variation of the efficiency of the engine cycle with the NC parameter of the system with non-commutative HO with relativistic and GUP effect as the working substance. The red solid line is the efficiency of the harmonic oscillator.}
\label{Fig4}
\end{figure}

We have considered $T_1 = 2K$, $T_2 = 1K$, $\omega = 4$ and $\omega^{'} = 3$ for the evaluation the efficiency of the Stirling engine (shown in Fig.~(\ref{Fig4})) with the variation of the NC parameter. We numerically analyze the efficiency of the engine. There is a steep increase in the efficiency of the Stirling cycle with the variation of the NC parameter. So, for this working medium with relativistic and GUP correction, we encounter a catalytic effect in the efficiency of the cycle with the increase of the NC parameter. The range of the NC parameter is considered in such a way that it does not exceed the extreme quantum gravity limit~\cite{castro}. Though it seems like that the efficiency of the engine is monotonously increasing with the NC parameter, there is a bound on it due to the accessible range of the NC parameter. Following the methodology proposed in~\cite{sdey,mkho}, one can have the experimental realization of the engine models in NC spacetime with a harmonic oscillator as the working medium. We can infer from the plot shown in Fig.~(\ref{Fig4}) that the non-commutative parameter along with the relativistic correction has an impact on the efficiency of the engine. It is a surprising fact that this model when considered as the working medium for the same engine we can visualize the effect of both the relativistic as well as the non-commutative parameter. The model dependency of the non-commutative parameter is confirmed from this analysis.


\section{Conclusion and Discussion}\label{sec5}
The non-commutative harmonic oscillator with relativistic and GUP correction as a working medium for the Stirling cycle, out-performs the harmonic oscillator as a working medium in terms of the efficiency of the cycle. With the increase in the NC parameter, the efficiency of the engine gets a boost. We have considered the range of the NC parameter in such a way that it is bounded by the extreme quantum gravity limit. We encounter a catalytic affect in the efficiency of the cycle with the harmonic oscillator as the working substance. But we get a constant efficiency with 1-D infinite potential well as the working medium for the same cycle. So, we can infer that the harmonic oscillator model is an effectual working substance relativistic NC spacetime over the 1-D infinite potential well model. The effect of non-linearity that we encounter in the Hamiltonian appears due to the NC parameter along with the GUP correction. This requires energy for the implementation in the thermodynamic cycles.

 In different application areas of the quantum theory~\cite{mhu}, the relativistic NC spacetime can be a better resource. This is an open area for exploration. In previous works~\cite{tzh,jwan,gho}, coupled working medium has been analyzed. One can utilize the NC spacetime for the analysis of coupled working medium, and it can be further extended for the exploration of non-Markovian reservoirs in relativistic NC spacetime. Here, in our work, we have analyzed the heat cycle. For the generic statement about the boost on efficiency for the engine cycle in NC spacetime, further investigation of the existing heat cycles, are required. Even one can study the nature of the coefficient of performance for the refrigerator cycles using the different spacetime models. It will be fascinating to study the effect of the NC spacetime even in the irreversible cycles~\cite{liu1,dalk} and how they affect the quantum phase transition.

Being motivated by works~\cite{chatt,chatt1,chatt2}, generalized uncertainty principle can be used to develop a bound in the efficiency of different thermodynamic variables, and thereafter, in various thermodynamic processes and thermodynamic cycles. The experimental models proposed for NC spacetime~\cite{sdey,mkho} have provided a boost for the exploration of NC spacetime in different application areas. Being motivated by this works, one can design experimental models for the analysis of the thermodynamic process and thermodynamic cycles in NC spacetime. In previous works on experimental analysis~\cite{deyyy}, it is shown that the non-commutative systems are more entangled than the usual quantum systems. The experimental development in this direction will help us to analyze the effect of non-commutative systems in the entanglement property and even to the thermodynamic processes.

\section{Acknowledgement}
The authors gratefully acknowledge  for the useful discussions and suggestions from Dr. Raam Uzdin, Senior Lecturer (Assist. Prof.) at the Hebrew University of Jerusalem, Faculty of Science, Fritz Haber Center for Molecular Dynamics Institute of Chemistry.


\bibliographystyle{h-physrev4}



\end{document}